\documentclass[12pt]{article}
\begin{document}
\sf

\begin{center}
   \vskip 2em
  {\LARGE \sf Yang-Mills theory  \`{a} la string}
\vskip 3em
  {\large \sf Riccardo  Capovilla${}^{(1)}$ and
Jemal Guven${}^{(2)}$  \\[2em]}
\em{
 ${}^{(1)}$ Departamento de F\'{\i}sica \\
 Centro de Investigaci\'on y de Estudios Avanzados\\
Apdo Postal 14-740, 07000 M\'exico,
D. F., MEXICO \\[1em]
${}^{(2)}$ Instituto de Ciencias Nucleares\\
 Universidad Nacional Aut\'onoma de M\'exico\\
 Apdo. Postal 70-543, 04510 M\'exico, D.F,, MEXICO}
\end{center}
 \vskip 1em
 \vskip 1em

\begin{abstract}
\sf
A surface of codimension higher than one embedded in an ambient space
possesses a connection associated with the rotational freedom of
its normal vector fields. We examine the Yang-Mills functional associated
with this connection. The theory it defines differs from Yang-Mills theory
in that it is a theory of surfaces. We focus, in particular,
on the Euler-Lagrange equations describing this surface, introducing a
framework which throws light on their relationship to the Yang-Mills equations.
\end{abstract}

\today
\vskip 1em


\vskip 3em

{\sl Dedicated to Octavio Obreg\'on, on the occasion of his 60th birthday.}

\vskip 3em

Consider a surface of codimension $N$ embedded in an ambient space.
There are then $N$ normal vector fields. Let us suppose that $N$ is
two or higher. The normal vectors are then defined only up to a
rotation, and a sign. There is a natural $O(N)$ connection on the
surface associated with this freedom known as the {\it normal}
connection, or extrinsic twist. This connection possesses a
curvature. While the role it plays is very different from that
played by the extrinsic curvature of the surface, the {\it normal}
curvature  is not independent of the latter: the Ricci integrability
conditions determine it completely in terms of a quadratic in the
extrinsic curvature \cite{Spivak}.

There are various interesting local geometrically invariant
functionals one can construct out of the normal curvature. Several
of these characterize geometrical features peculiar to particular
dimensions: one such invariant, characterizing two-dimensional
surfaces embedded in four-dimensions, is the integral of the normal
curvature itself; this invariant was introduced by Polyakov in the
context of a stringy description of QCD \cite{Polyakov}. For
arbitrary surface dimensions, there is a natural invariant quadratic
in the normal curvature: the $O(N)$ Yang-Mills functional.
Appearances are deceptive however; this is not a genuine Yang-Mills
theory. The dynamical variables are the embedding functions of the
surface, not the connection itself.  More appropriately, one should
consider the theory it defines as an induced Yang-Mills theory. This
is the functional we will focus on in this note. In particular, we
would like to clarify the relationship between the two theories.
This situation is analogous to that for Regge and Teitelboim's model
for gravity in terms of an Einstein-Hilbert action, where the
embedding functions, rather than the metric, appear as the dynamical
variables \cite{RT}. To spell out the parallel, we have adapted the
title of their paper -- {\it Gravity \`{a} la string} -- to its
Yang-Mills counterpart.

In particular, if the surface is four-dimensional this functional is
also conformally invariant. It is one of the few low order conformal
invariants of the surface geometry. Perhaps the best known among
these invariants is quadratic in the Weyl tensor, which depends only
on the intrinsic geometry\cite{Wald}. There are also polynomial
invariants --- apparently not so well-known to physicists
--- associated with the extrinsic geometry of the surface
\cite{Willmore}. The simplest invariants of this kind are
constructed using the traceless part of the extrinsic curvature.
There are two independent quartics of this form \cite{CCG}. Indeed,
using the Gauss-Codazzi equations for the surface, it is possible to
express the Weyl invariant as a linear combination of the two. In
addition, a third invariant can be associated with a non-trivial
competition between a quartic in extrinsic curvature and a quadratic
in gradients, neither of which alone is conformally invariant
\cite{guven}.

\vskip 3em
Consider a $d$-dimensional surface $\Sigma$
embedded in $R^{d+N}$, where $N\ge 2$,  described parametrically
as follows:
\begin{equation}
x = X (\xi^a)\,.
\label{eq:xX0}
\end{equation}
Here $x=(x^1,\dots,x^{d+N})$ are local coordinates for the ambient
space, $\xi^a$, $a=1,\dots, d$ are $d$ arbitrary coordinates on the
surface, and $X = (X^1, \dots, X^{d+N}) $ denote the embedding
functions. The only geometrically significant derivatives of $X$ are
those encoded in the metric tensor $g_{ab}= e_a\cdot e_{b}$ and the
extrinsic curvature tensor $K_{ab}{}^I= e_a\cdot \partial_b \,n^I
=K_{ba}{}^I$, where $e_a= \partial_a X $ are the $d$  coordinate
tangent vectors, and $n^I$ are any $N$ mutually orthogonal normal
vectors ($I=1,\cdots, N$). The Gauss-Weingarten equations describing
how this surface gets embedded in the ambient space are given by
($\nabla_a$ is the usual metric compatible covariant derivative on
$\Sigma$)
\begin{eqnarray}
\nabla_a e_b &=& - K_{ab}{}^I n_I \,, \label{eq:gw1}\\
\tilde \nabla_a n^I &=&  K_{ab}{}^{I}  g^{bc} e_c \,. \label{eq:gw2}
\end{eqnarray}
We have introduced the $O(N)$ covariant derivative on the surface,
$\tilde \nabla_a$, associated with its invariance under rotations of
the normal vectors:
\begin{equation}
\tilde \nabla_a \Phi^I = \partial_a \Phi^I + A_a{}^{I}{}_J \, \Phi^J\,,
\end{equation}
with $\Phi^I$ an arbitrary normal scalar, and
where the normal connection is given by
\begin{equation}
A_a{}^{IJ} =   n^I \cdot \partial_a n^J = -  A_a{}^{JI}\,,
\end{equation}
of course  for $N=1$ it vanishes identically.
We denote by
$F_{ab}{}^{IJ}$ the normal curvature associated with the  normal connection
$A_a{}^{IJ}$: $[\tilde\nabla_a,\tilde\nabla_b] \Phi^I = F_{ab}{}^{IJ}\Phi_J$.
Explicitly:
\begin{equation}
F_{ab}{}^{IJ} = \partial_a A_b{}^{IJ} +
A_a{}^{IK} A_{b\,K}{}^J - (a \, \leftrightarrow b)\,.
\label{eq:curva}
\end{equation}
There are integrability conditions associated with the
Gauss-Weingarten  equations. Besides the well-known
Gauss-Codazzi-Mainardi integrability conditions,  when the number of
extra dimensions is two or higher, one has the Ricci identities
\cite{Spivak}:
\begin{equation}
F_{ab}{}^{IJ} = K_{ac}{}^I K^c{}_b{}^J - (I \,\leftrightarrow\,  J)\,.
\label{eq:Ricci}
\end{equation}
Thus the normal curvature associated with $A_a{}^{IJ}$ is completely
determined by the induced metric $g_{ab}$ and the extrinsic
curvature $K_{ab}{}^I$.

With the normal curvature, we can construct the geometric functional
($dV $ is the surface volume element constructed with $g_{ab}$)
\begin{equation}
I_0 [X] = {1\over 4} \int dV \, F^{ab}{}_{IJ} F_{ab}{}^{IJ}\,.
\label{eq:YM}
\end{equation}
Superficially, it resembles  an $O(N)$ Yang-Mills theory on the
surface. It differs in the important respect that $I_0$ is a
functional of the embedding functions  $X$ and not of the connection
$A_a{}^{IJ}$. One important consequence of this fact is that, while
this functional is of first order in derivatives of the connection
$A_a{}^{IJ}$, it is of second order in derivatives of the embedding
functions $X$; this can be seen by looking at the Ricci identities
(\ref{eq:Ricci}): $F^2 \sim K^4$, it is quartic in powers of the
extrinsic curvature. We remark that, if $d=4$, the action $I_0$ is
conformally invariant.

It is instructive to examine how the  difference
in the choice of the field variables manifests
itself at the level of the Euler-Lagrange equations.
We will compare the Euler-Lagrange derivative ${\cal E}$,
determining the response of the functional $I_0$ to a
surface deformation, $X \to X + \delta X$,
\begin{equation}
\delta_X I_0 = \int dV \, {\cal E}  \cdot \delta X\,,
\end{equation}
with the Euler-Lagrange derivative ${\cal E}^{a}{}_{IJ}$ corresponding to
a variation in the connection $A_a{}^{IJ}  \to A_a{}^{IJ} + \delta A_a{}^{IJ}$,
\begin{equation}
\delta_A I_0 = \int dV \, {\cal E}^{a}{}_{IJ}  \;\delta A_a{}^{IJ}\,.
\end{equation}
The first obvious thing to note is that the two differ in the number
of independent variations: $d+N$ in the former versus $d\times
N(N-1)/2$ in the latter. For the latter, using the expression
$\delta_A F_{ab}{}^{IJ} = 2 \nabla_{[a} \delta A_{b]}{}^{IJ}$, for
the variation of the normal curvature, we have the Yang-Mills
equations
\begin{equation}
{\cal E}^a{}_{IJ} = \nabla_b \; F^{ab}{}_{IJ} = 0\,.
\label{eq:YME}
\end{equation}

In order to obtain the variation of $I_0$ with respect to $X$,
various strategies are possible. One could vary the connection
$A_a{}^{IJ}$ directly, using the variational expressions obtained
{\it e.g.} in  Ref. \cite{defos}. This approach, however, has the
disadvantage of introducing a connection deformation which plays no
role at the end, but that appears annoyingly in intermediate
calculation. A second approach would be to use the Ricci identities
(\ref{eq:Ricci}) to express the normal curvature in terms of the
metric and the extrinsic curvature, and then vary these geometrical
quantities. However, in this way the connection with Yang-Mills
theory is blurred. The strategy we adopt is  to introduce auxiliary
variables, along the lines first suggested in Ref. \cite{auxil} in a
different context. We thus construct the functional
\begin{eqnarray}
I[ X,& e_a , n^I , A_a{}^{IJ} , g_{ab} , \lambda_a{}^{IJ} ,
\lambda^{ab} , \lambda^{IJ}, \lambda^a{}_I , {\cal F}^a ] = I_0 [
A_a{}^{IJ}, g_{ab} ]- \int dV \,\left[  \lambda_{IJ}^a \,(A^{IJ}_a -
n^I \cdot \nabla_a n^J) \right. \nonumber \\  &- \left. {1\over 2}
\lambda^{ab} (g_{ab} - e_a \cdot e_b) - {1\over 2} \lambda_{IJ}
(n^I\cdot n^J - \delta^{IJ}) -  \lambda^a{}_I (n^I\cdot e_a) + {\cal
F}^a\cdot (e_a- \partial_a X) \right]\,. \label{eq:aux}
\end{eqnarray}
Following the approach introduced in Ref. \cite{auxil}, we treat the
connection $A_a{}^{IJ}$ as variables independent of $X$. We must then
introduce Lagrange multipliers to enforce the constraints that
connect $A_a{}^{IJ}$ to the geometry. In this approach, the induced
metric $g_{ab}$, as well as the basis vectors $\{ e_a , n^I \}$, are
also treated as independent. The innocuous apearing constraints
anchoring the tangent vectors to a derivatives of the embedding
functions $X$, via the Lagrange multipliers ${\cal F}^a$ will play a
very important role: they identify the conserved momentum. At first
sight, the introduction of a plethora of auxiliary variables appears
only to complicate matters. However, as we will see, it is downhill
from here: the inplementation of the constraints is straightforward.
In particular, we do not need to know the variation of any of the
surface geometric tensors in terms of $\delta X$. The geometrical
significance of the different Lagrange multipliers will emerge
automatically.

We start by considering the variation of (\ref{eq:aux}) with respect
to $A^{IJ}_a$. We obtain immediately
\begin{equation}
\tilde\nabla_b F^{ab}{}_{IJ} = \lambda^a{}_{IJ}\,.
\end{equation}
Thus the Lagrange multiplier $\lambda^a{}_{IJ}$ is identified with
the Euler-Lagrange derivative ${\cal E}^ a{}_{IJ}$ introduced above
(see Eq. (\ref{eq:YME})), and it
 acts as a source for the
Yang-Mills field.
It is clear that, if $\lambda^{IJ}_a\ne 0$,
the Yang-Mills equations are not satisfied.

Next, consider variations with respect to the tangent vectors $e_a$:
\begin{equation}
{\cal F}^a = -  \lambda^{ab} e_b + \lambda^a{}_I n^I  \,.
\end{equation}
This is simply an expansion for ${\cal F}^a$ in terms of the
basis adapted to the surface $\{e_a , n^I \}$.

The equations for the embedding functions $X$ is the statement
that ${\cal F}^a$ is covariantly conserved:
\begin{equation}
\nabla_a {\cal F}^a =0\,.
\label{eq:conserv}
\end{equation}
By examining the translational invariance of the functional  $I$,
it is clear that ${\cal F}^a$ is the conserved momentum  density. Under
$X\to X+a$, where $a$ is constant, we have $I\to I + \delta I$, where
\[
\delta I = - a \cdot \int dV\, \nabla_a {\cal F}^a\,.
\]
In particular, note that we can decompose the conservation law
(\ref{eq:conserv}) into its tangential and normal parts by
projection and using the Gauss-Weingarten equations (\ref{eq:gw1}),
(\ref{eq:gw2}):
\begin{eqnarray}
-\nabla_a \lambda^{ab} + \lambda^a{}_I \; K_a{}^{b\, I} &=& 0\,, \label{eq:const}\\
\nabla_a \lambda^a{}_I + \lambda^{ab} K_{ab\, I} &=& 0 \label{eq:consn}\,.
\end{eqnarray}
The tangential projection is the Bianchi identity associated with
the reparametrization invariance of the functional $I$. The second,
normal, is the Euler-Lagrange derivative for the functional  $I_0 [
X ]$ with respect to the embedding functions $X$ (see {\it. e.g.}
\cite{ACG}). What remains to be done is to identify the geometric
content of the components $\lambda^{ab}$ and $\lambda^a{}_I$.  For
this, consider now the variation with respect to the induced  metric
$g_{ab}$. It identifies the tangential component of the stress
tensor $\lambda^{ab}$ with the metric stress tensor for the
Yang-Mills functional (\ref{eq:YM}):
\begin{equation}
 \lambda^{ab}  = T^{ab} =
F^{ac}{}^{IJ} F^b{}_c{}_{IJ} - {1 \over 4} g^{ab} F_{cd}{}^{IJ}
F^{cd}{}_{IJ}\,, \label{eq:tab}
\end{equation}
where the metric stress tensor $T^{ab}$ is defined in the usual way as
\begin{equation}
T^{ab}  =  {1 \over 2 \sqrt{g}} { \delta I_0 \over \delta g_{ab}}\,.
\label{eq:tdef}
\end{equation}
Note that $T^{ab}$ itself is not conserved. If $d=4$, it is,
however, traceless, the hallmark of conformal invariance. We find
that the tangential projection  of the conservation law for ${\cal
F}^a$ (\ref{eq:const}) now reads
\begin{equation}
\nabla_a T^{ab} = K_a{}^{b\, I}  \;  \lambda^a{}_I\,.
\end{equation}
The extrinsic curvature of the surface and the normal component of the momentum
density
provide a source for
the divergence of the metric stress tensor $T^{ab}$.

To determine the normal component of ${\cal F}^a$, $\lambda^a{}_I$, we consider
the last variation, with respect to   the normal vectors $n^I$:
\begin{equation}
\lambda^a{}_{IJ} \nabla_a n^J + \nabla_a (\lambda_{IJ}^a n^J) +
 \lambda_{IJ} n^J + \lambda^a{}_I e_a =0\,.
\end{equation}
The tangential projection of this equation identifies the Lagrange
multiplier $\lambda^a{}_I$, or the normal component of the momentum
density, for us:
\begin{equation}
\lambda^a{}_I  =  - 2   \lambda^b{}_{IJ} K^a{}_b{}^J  =  - 2   {\cal
E}^b{}_{IJ} K^a{}_b{}^J  = -2   (\tilde\nabla_c F^{bc}{}_{IJ})
K^a{}_b{}^J\,.
\end{equation}
The normal projection identifies the remaining Lagrange multiplier
$\lambda_{IJ}$ as
\begin{equation}
\lambda_{IJ} = 2 A_{a(I}{}^K \lambda^a{}_{|K|J)}\,,
\end{equation}
together with the vanishing of the divergence of the Lagrange multiplier
$\lambda^a{}_{IJ}$, or equivalently, of the Yang-Mills Euler-Lagrange derivative:
\begin{equation}
\tilde\nabla_a \lambda^a{}_{IJ} = \nabla_a {\cal E}^a{}_{IJ} = 0\,.
\end{equation}
This is the Bianchi identity associated with the $O(N)$ invariance of the
Yang-Mills functional.

We conclude that the momentum density takes the form
\begin{eqnarray}
{\cal F}^a &=&  -  T^{ab} e_b - 2
K^{a}{}_{b}{}^J \; {\cal E}^b{}_{IJ}
 n^I\,,  \nonumber \\
&=&-  T^{ab} e_b - 2 K^{a}{}_{b}{}^J \; (\nabla_c F^{cb}{}_{IJ} )
 n^I\,.
\end{eqnarray}
The first line is valid for {\it any} functional of the normal
connection. It requires only to determine the variations of the
functional with respect to the induced metric $g_{ab}$, to obtain
$T^{ab}$ via Eq. (\ref{eq:tdef}), and with respect to the connection
to obtain ${\cal E}^a{}_{IJ}$. The second line is specific to the
Yang-Mills model and $T^{ab}$ is given by (\ref{eq:tab}). Note that,
unless the Yang-Mills equations hold, there is a non-trivial normal
component of the momentum density.

It is immediate to see that the tangential projection of the
divergence of ${\cal F}^a$, (\ref{eq:const}), is an identity. The
vanishing of the normal projection of the divergence of ${\cal F}^a$
produces the Euler-Lagrange equations of the functional $I$ with
respect to $X$ (see {\it e.g.} \cite{ACG}):
\begin{equation}
2 \tilde\nabla_a [ K^{a}{}_{b}{}^J \; (\nabla_c F^{cb}{}_{IJ} ) ] -
K_{ab}{}^I T^{ab}=0\,. \label{eq:eom}
\end{equation}
These equations are fourth order in derivatives of the embedding
functions. Even if the Yang-Mills equations are satisfied, so that
the first term vanishes, one must still contend with the second
term. In this case, however,  the equations reduce to second order:
curiously, they are reminiscent of the Euler-Lagrange equations for the
Regge-Teitelboim model \cite{RT}.

In conclusion, we have described a theory of embedded surfaces
described by a Yang-Mills functional. In particular, we have
examined the relationship between this theory and Yang-Mills theory
at the level of the Euler-Lagrange equations as well as the
conserved momentum. While the two theories differ, there are also
intriguing connections. A detailed discussion will be presented
elsewhere.

\newpage

\noindent{\large  Acknowledgments}
\vspace{.5cm}

Partial support from CONACyT grants 44974-F, 51111 as well as DGAPA PAPIIT
grant IN119206-3 is acknowledged.


\begin{thebibliography}{99}


\bibitem{Spivak} M. Spivak,  {\it A Comprehensive Introduction to
Differential Geometry. Vol.Four, Second Edition} (Publish or
Perish, 1979).

\bibitem{Polyakov} A. Polyakov, {\it Nucl. Phys. B} {\bf 268} 406 (1986).

\bibitem{RT} T. Regge and C. Teitelboim, {\it Proceedings
of the Marcel Grossman Meeting}, Trieste, Italy, (1975),
ed.  Ruffini R (North-Holland, Amsterdam, 1977) 77.


\bibitem{Wald} R. M. Wald, {\it General Relativity}
(University of Chicago Press 1984).

\bibitem{Willmore} T.J. Willmore, {\sl Total Curvature in Riemannian
Geometry} (Chichester: Ellis Horwood, 1982).

\bibitem{CCG} R. Capovilla, R. Cordero, J. Guven, {\it Mod. Phys. Lett. A} {\bf 11}
2755 (1996).

\bibitem{guven} J. Guven {\it J. Phys. A: Math. and Gen.} {\bf 38} 7943 (2005).

\bibitem{defos} R. Capovilla and J. Guven, {\it Phys. Rev. D} {\bf 51} 6736 (1995).

\bibitem{auxil} J. Guven,
{\it J. Phys. A: Math and Gen.} {\bf 37}
 L313 (2004).

\bibitem{ACG} G. Arreaga, R. Capovilla, and J. Guven,
{\it Ann. Phys.} {\bf 279} 126 (2000).


\end{thebibliography}
\end{document}